\documentclass[a4paper,12pt]{article}
\usepackage[utf8]{inputenc}
\usepackage{amsmath}
\usepackage{amssymb}
\usepackage{slashed, color}
\title{Short Distance  Modification of the Quantum  Virial Theorem}
\author{Qin Zhao$^{1}$, Mir Faizal$^{2,3}$,  Zaid Zaz$^{4,5}$\\\\
$^1$ Department of Physics, National University of Singapore, \\
 2 Science Drive 3, Singapore\\
$^2$ Irving K. Barber School of Arts and Sciences, \\ University of British Columbia,
 \\ Kelowna, British Columbia, V1V 1V7, Canada.  \\$^3$ Department of Physics and Astronomy,\\
University of Lethbridge, \\
Lethbridge,   Alberta, T1K 3M4, Canada. \\ 
 $^4$ Theoritical Physics Division,  Department of Physics, \\ National Institute of Technology, \\
Srinagar, Kashmir-190006, India 
\\$^5$ Department of Electronics and Communication Engineering, \\University of Kashmir, 
Srinagar, Kashmir-190006, India\\}
\date{}
\begin{document}

\maketitle

\begin{abstract}
In this letter, we will analyse the deformation of a semi-classical 
gravitational system from minimal measurable length scale. 
In the semi-classical approximation, the gravitational field will be analysed as 
a classical field, and the matter fields will be treated quantum mechanically. 
Thus, using this approximation, this system will be represented by a 
  deformation of  Schr\"odinger-Newton equation by the generalised uncertainty principle 
  (GUP).
We will analyse the effects of this  GUP deformed Schr\"odinger-Newton equation
on the behaviour of  such a semi-classical  gravitational system.  
As the quantum mechanical virial theorem can be 
obtained using the Schr\"odinger-Newton equation, 
a short distance modification of the Schr\"odinger-Newton equation will also 
result in a short distance modification of the quantum mechanical virial theorem. 
\end{abstract}

Even though there have been various attempts to formulate a quantum theory of gravity, 
there are serious problems with obtaining a fully quantum theory of gravity. 
This has motivated the study of semi-classical quantum gravity, and in such approaches 
the gravitational field is treated classically and the matter fields are treated quantum 
mechanically. Thus, in this approximation, the gravitational field 
is  a classical field, but
it is sourced by quantum mechanical matter fields, $
G_{\mu\nu} =  {8\pi G} \langle \psi |\hat{T}_{\mu\nu}|\psi \rangle/{c^2}, 
$
where $G_{\mu\nu}$ is the  Einstein tensor for a   classical space-time, 
$\hat{T}_{\mu\nu}$ is the  operator for the energy-momentum tensor, 
and $|\psi\rangle $ is the wave-function of the matter fields.  
Furthermore,  the Newtonian gravity has been observed to be the 
correct approximation to 
general relativity till the smallest length scale to which general relativity 
has been tested \cite{mz4}. 
Thus, it is possible to use the Newtonian gravitational field in this approximation, 
and 
describe this system using a Schr\"odinger-Newton equation \cite{1newt}-\cite{newt0}, 
\begin{equation}\label{action_normal}
i\hbar \frac{\partial\psi}{\partial t} 
= H \psi = \frac{1}{2 m}\hat{P}^2\psi + m\Phi(\textbf{R}, t)\psi,
\end{equation}
where $H$ is the Hamiltonian operator and $\Phi(\textbf{R}, t)$ 
is the Newtonian  potential, 
and the mass density of the system is $m|\psi(\textbf{R}, t)|^2 
= \rho(\textbf{R}, t)$.  
It has been demonstrated that such a Schr\"odinger-Newton equation 
can be obtained from a non-relativistic limit of 
self-gravitating Klein-Gordon and Dirac fields \cite{rela12}. 
The Schr\"odinger-Newton equation has been used to study the semi-classical 
behaviour of various interesting gravitational systems \cite{equation}-\cite{equation1}. 
It has been proposed that such a  Schr\"odinger-Newton equation  can be tested experimentally 
using an optomechanical systems  \cite{Gan:2015cxz}-\cite{Grossardt:2015moa}. 
However, it has also been demonstrated that 
optomechanical systems can be used to test the deformation 
of a quantum mechanical system by the generalised uncertainty principle 
\cite{Pikovski:2011zk}.  
Thus, it becomes important to study the deformation of Schr\"odinger-Newton 
equation by generalised uncertainty principle (GUP), as this deformation
can be tested experimentally 
using optomechanical systems. In fact,  as
the general relativity has not been tested below  $0.4  mm$  \cite{mz4}, 
 it is possible that GUP can produce  interesting short distance  modification 
 for a gravitational system. This modification can alter the behaviour of the system 
 and can be detected using optomechanical systems. Thus, it is very interesting to 
 analyse the GUP deformation of  Schr\"odinger-Newton equation, and so such 
 deformation of  Schr\"odinger-Newton equation has been thoroughly studied
 \cite{gravity0}-\cite{garvitya}.

Furthermore, there is another reason why such a the GUP deformation of 
Schr\"odinger-Newton equation
would be interesting. The GUP deformation of the Schr\"odinger-Newton equation can be 
viewed as an intermediate step between 
the full quantum theory of gravity and semi-classical quantum gravity. 
This is because  a  universal feature of almost all approaches to quantum gravity 
is the existence of a minimum measurable length scale.
It is postulated that physical measurements below this scale are not possible. 
In string theory, which is one of the most popular approaches to quantum gravity, 
the smallest scale at which  space-time can be probed is the string length scale. 
This is because strings are the smallest probes that exist in perturbative string theory, 
and so it is not possible to probe space-time below string length scale
\cite{z2}-\cite{2z}. Even  in loop quantum gravity,  the existence of a minimum length
has a very interesting consequence
as it turns the big bang into a big bounce \cite{z1}.
Moreover arguments from black hole physics also suggest that any theory of quantum gravity 
 must be equipped with a minimum length scale \cite{z4}-\cite{z5}. 
 This is due to the fact that the energy required to probe any region of space below 
 the Plank length 
is greater than the energy required to create a mini black hole in that region of space.
Thus, by incorporating an intrinsic minimal length scale in the Schr\"odinger-Newton equation, 
certain features of the quantum gravity can be analysed beyond the purely semi-classical 
approximation. 

This minimal length can be incorporated into such a system by studding the GUP deforming 
of the Schr\"odinger-Newton equation \cite{gravity0}-\cite{garvitya}.
The GUP deformation of quantum mechanical systems is motivated from the fact that 
usual Heisenberg uncertainty principle is not 
consistent with the existence of a minimum length scale in the system.  Thus, 
the Heisenberg uncertainty principle  has to be generalised to a generalised uncertainty 
principle  (GUP) to incorporate the existence of such a minimum measurable 
length scale \cite{1}-\cite{15}. 
It is known that the uncertainty principle is related to   the Heisenberg algebra, and so 
any modification of the uncertainty principle will deform the  
Heisenberg algebra \cite{17}-\cite{53}. The deformation of the Heisenberg algebra will 
in turn modify the representation of the momentum operator in a position basis \cite{18}-\cite{10}. 
These effects can be measured from the 
deformation of the  Heisenberg algebra to \cite{18}-\cite{10}
\begin{equation}\label{GUP}
[\hat{X}_i, \hat{P}_j] = i\hbar(\delta_{ij} + \beta\delta_{ij}\hat{P}^2 + 2\beta \hat{P}_i \hat{P}_j).
\end{equation}
It can be demonstrated using the Jacobi identity that the other commutators for 
this system do not get deformed \cite{54}-\cite{51}
\begin{equation}
[\hat{X}_i, \hat{X}_j] = 0 = [\hat{P}_i, \hat{P}_j]. 
\end{equation}
It may be noted that this deformation occurs due to the existence 
of an extended structure (minimum length) in space-time,
and hence it also deforms the   standard Poisson brackets as
$\{x, p\} = 1+ \beta p^2 $ \cite{gravity01}.  
This is the minimal extension of the Heisenberg algebra, and to analyse it further, we  define   
\begin{align}
\begin{aligned}
& \hat{X}_i\psi(\textbf{x}) = \hat{x}_i \psi(\textbf{x}) \\
& \hat{P}_i\psi(\textbf{x}) = \hat{p}_i(1+\beta \hat{p}^2)\psi(\textbf{x}),
\end{aligned}
\end{align}
where $\hat{x}_i$ and $\hat{p}_j = -i\hbar\frac{\partial}{\partial {\hat{x}}_j}$ satisfy the canonical commutation relations $[\hat{x}_i, \hat{p}_j] = i\hbar\delta_{ij}$. 
Then, it is easy to show that in the first order of $\beta$, 
\eqref{GUP} is satisfied by this deformed momentum operator  
(here we neglect terms of order $\beta^2$ and higher). 
Physically, we interpret $\hat{p}_i$ as the momentum in 
the IR limit and $\hat{P}_j$ as
momentum in the UV limit of the 
theory. 
It is worth mentioning here that  the minimum measurable 
length has to exist at least at the Planck scale. 
However, it can be argued that the minimum measurable length  can exist
at a much greater scale than the Planck scale. 
If the generalised uncertainty principle deforms all field theoretical systems 
in the same way, then the 
bound on the existence of the minimum measurable length scale is the electroweak scale
\cite{54}. Even this can have low energy consequences. 
In fact,  such corrections have been obtained  for various physical phenomena 
such as Lamb shift and  Landau levels  \cite{mz1}.
Now we can assume that there is a minimum measurable length scale in the
background space-time, at a scale much larger than Planck scale. This will deform the 
Heisenberg uncertainty principle to GUP which will have low energy effects.

Now by substituting $\hat{X}_i, \hat{P}_i$ by $x_i, p_i$, 
we can rewrite the Schr\"odinger-Newton  equation  \eqref{action_normal}  
as (to the leading order in $\beta$)
\begin{equation}\label{action_gup}
i\hbar \frac{\partial\psi}{\partial t} = H \psi = \frac{1}{2 m} \hat{p}^2\psi
+ \frac{\beta}{ m} \hat{p}^4\psi + m\Phi(\textbf{r}, t)\psi. 
\end{equation}
We would like to point out that the GUP deformation of such a system has already been studied 
 \cite{gravity0}-\cite{garvitya}.  However, we shall use the GUP deformation of the 
 quantum mechanical   virial theorem for such a Schr\"odinger-Newton  equation. 
 The quantum mechanical virial theory has been recently obtained using a
 Schr\"odinger-Newton  equation, and it has been demonstrated 
 that such a quantum mechanical  virial theory can have interesting cosmological
 consequences \cite{witten}. So, in this letter, we will analyse the GUP deformation of such a 
 quantum mechanical virial theory using a GUP deformed Schr\"odinger-Newton  equation. Now the 
 moment of inertia for this system is $I = \frac{1}{2}m\int d\textbf{r} r^2|\psi|^2$, and so we 
 can write 
\begin{align}
\begin{aligned}
\dot{I} & = -\frac{im}{2\hbar}\int d\textbf{r} r^2(\psi^*H\psi - \psi H\psi^*) \\
& = -\frac{im}{2\hbar}\int d\textbf{r} \psi^*[r^2, H]\psi \\
& = \frac{i\hbar}{4}\int d\textbf{r}\psi^*[r^2, \nabla^2 - 2\beta \hbar^2 \nabla^4]\psi,
\end{aligned}
\end{align}
noting that $[r^2, \Phi]=0$. 
Using the derivative along time direction,  we can obtain
\begin{align}
\begin{aligned}
\ddot{I} & = \frac{1}{4}\int d\textbf{r}\psi^*[[r^2, \nabla^2 - 2\beta \hbar^2 \nabla^4], H]\psi.
\end{aligned}
\end{align}
Using this result, it can be demonstrated that the following result holds for this system, 
\begin{align}
\begin{aligned}
&& [[r^2, \nabla^2], -\frac{\hbar^2}{2m}\nabla^2 + m\Phi] = - 4m(\textbf{r}\cdot \nabla\Phi) - \frac{4\hbar^2}{m}\nabla^2, \\
\end{aligned} \\
\begin{aligned}
& [[r^2, \nabla^2], \frac{\hbar^4\beta}{m}\nabla^4] + [[r^2, -2\beta\hbar^2\nabla^4], -\frac{\hbar^2}{2m}\nabla^2] \\
&=\frac{2\hbar^4\beta}{m}[[r^2, \nabla^2], \nabla^4] \\
&= \frac{2\hbar^4\beta}{m} [r^2\nabla^6 - \nabla^2 r^2\nabla^4 - \nabla^4 r^2\nabla^2 +\nabla^6 r^2]
\\
& = \frac{2\hbar^4\beta}{m} * 16 \nabla^4,
\end{aligned}
\end{align}
where we have $ 
\nabla^2 r^2\nabla^4 = 6 \nabla^4 + 4\textbf{r}\cdot\nabla \nabla^4 + r^2\nabla^6, $ and $ \nabla^4 r^2\nabla^2 = 20 \nabla^4 + 8\textbf{r}\cdot\nabla \nabla^4  
+ r^2\nabla^6,$ along with $
\nabla^6 r^2 = 42\nabla^4 + 12 \textbf{r}\cdot\nabla \nabla^4 +  r^2\nabla^6$. We also obtain 
\begin{align}
\begin{aligned}
\quad [[r^2, -2\beta\hbar^2\nabla^4],   m\Phi]
& = 20 \hbar^2 \beta m( \nabla^2 \Phi) + 8\hbar^2 \beta m (\textbf{r}\cdot \nabla \nabla^2\Phi) \\ 
& + 8\hbar^2\beta m(\nabla^2 \Phi)\textbf{r}\cdot  \nabla + 8\hbar^2\beta m(\textbf{r}\cdot \nabla\Phi) \nabla^2 \\
&+ 16\hbar^2\beta m(\textbf{r}\cdot  \nabla)\nabla\Phi\cdot\nabla.
\end{aligned}
\end{align}

Then, we can expand $\ddot{I}$ to 
\begin{align}
\begin{aligned}
\ddot{I} =& -\int d\textbf{r}\psi^*(m\psi (\textbf{r}\cdot\nabla\Phi) + \frac{\hbar^2}{m}\nabla^2\psi) +\int d\textbf{r}\psi^*\frac{8\hbar^4\beta}{m}\nabla^4\psi \\
& + \int d\textbf{r} 5\psi^* \hbar^2 \beta m( \nabla^2 \Phi)\psi + \int d\textbf{r} 2 \hbar^2\beta m \psi^* (\textbf{r}\cdot \nabla \nabla^2\Phi)\psi \\
&+ \int d\textbf{r} 2\hbar^2\beta m\psi^* (\nabla^2 \Phi)\textbf{r}\cdot  \nabla\psi + \int d\textbf{r} 2\hbar^2\beta m\psi^* (\textbf{r}\cdot \nabla \Phi) \nabla^2\psi \\
&+ \int d\textbf{r}\psi^* 4\hbar^2\beta m(\textbf{r}\cdot  \nabla)(\nabla\Phi\cdot\nabla\psi).
\end{aligned}
\end{align}
These results hold for any potential as we have not used the explicit form for 
the gravitational field. However, now we will use an explicit form for the 
gravitational field, and analyse this deformation of the Schr\"odinger-Newton equation
from GUP. It may be noted 
that if the GUP deforms all field theoretical systems equally, then 
the experimental bound on the GUP has to be of the order of electroweak scale.
However, there is no reason for the GUP to deform all field theoretical systems 
by the same amount, and thus we can assume that the value of the deformation parameter 
$\beta$ varies between different field theoretical systems. 
In this letter, we will analyse the GUP deformation of a gravitational system, 
and assume that the generalised uncertainty acts differently for gravitational systems. 
 As gravity has not been tested below  $0.4 mm$  \cite{mz4}, 
 such effects can be detected in future optomechanical experiments.  
So, now we  explicitly write the  form for $\Phi$, and analyse the GUP 
deformation of the Schr\"odinger-Newton equation 
\begin{equation}
\Phi = - G \int \frac{d\textbf{r}'\rho(\textbf{r}')}{|\textbf{r}-\textbf{r}'|}.
\end{equation}
Now we can use 
\begin{align}
\nabla\Phi = -G\int \frac{d\textbf{r}'\rho(\textbf{r}')(\textbf{r}-\textbf{r}')}{|\textbf{r}-\textbf{r}'|^3}, && \nabla^2\Phi =0.
\end{align}
Therefore, the terms containing $\nabla^2\Phi$ vanish\footnote{Here, we assume that we do not consider the singularity around $r=0$.}. Furthermore, it is also useful to take the wave-function in the form 
$\psi = \sqrt{\rho/m}\exp(i\theta)$ with $\theta$ real, and $\textbf{v}=\frac{\hbar\nabla\theta}{m}$, 
\begin{align}
\begin{aligned}
\ddot{I} & = -\int d\textbf{r} \rho (\textbf{r}\cdot\nabla\Phi) + \int d\textbf{r}\rho \textbf{v}^2 + \frac{\hbar^2}{m^2}\int d\textbf{r}|\nabla\sqrt{\rho}|^2  \\
& + \int d\textbf{r} \frac{8\hbar^4\beta}{m} (\nabla^2\sqrt{\frac{\rho}{m}} -\sqrt{\frac{\rho}{m}} \nabla\theta \cdot\nabla\theta)^2 \\
& + \int d\textbf{r} \frac{8\hbar^4\beta}{m} (\nabla\sqrt{\frac{\rho}{m}}\cdot\nabla\theta +\sqrt{\frac{\rho}{m}}\nabla^2\theta)^2 \\
& +\int d\textbf{r}2\hbar^2\beta m (\textbf{r}\cdot\nabla\Phi ) \psi^*\nabla^2\psi\\
&+ \int d\textbf{r}\psi^* 4\hbar^2\beta m(\textbf{r}\cdot  \nabla)(\nabla\Phi\cdot\nabla\psi).
\end{aligned}
\end{align}
This is the quantum virial theorem for the gravitational field deformed by the generalised uncertainty principle. It may be noted here 
we have taken the parameter $\beta$ bounded by short distance measurement of gravitational  potential  \cite{mz4}.  
In the limit case, where $\beta = 0$, then we just obtain the quantum virial theorem \cite{witten}
\begin{align}
\begin{aligned}
{\ddot{I}}_{\beta=0} & =  -\int d\textbf{r} \rho (\textbf{r}\cdot\nabla\Phi) + \int d\textbf{r}\rho v^2 + \frac{\hbar^2}{m^2}\int d\textbf{r}|\nabla\sqrt{\rho}|^2 \\
& = W + 2K  + 2Q,
\end{aligned}
\end{align}
where $K =\frac{1}{2} \int d\textbf{r}\rho \textbf{v}^2 $ is the  kinetic energy, and  
$Q$ is a new term and it is called as  the quantum energy \cite{witten}.
The potential of the system $W$ is given by 
\begin{eqnarray}
W &=& -\int d\textbf{r} \rho (\textbf{r}\cdot\nabla\Phi) \nonumber \\
&=& -G\int d\textbf{r} d\textbf{r}' \frac{\rho(\textbf{r})\rho(\textbf{r}')\textbf{r}\cdot(\textbf{r}-\textbf{r}')}{|\textbf{r}-\textbf{r}'|^3} \nonumber \\
&=& -\frac{G}{2}\int d\textbf{r}\frac{\rho(\textbf{r})\rho(\textbf{r}')}{|\textbf{r}-\textbf{r}'|}.
\end{eqnarray}
Then, in a steady state where $\ddot{I}=0$ and $K>=0$, we have  \cite{witten}
\begin{equation}
\frac{Q}{|W|} \leq \frac{1}{2}.
\end{equation}
It is easy to realise that when the phase 
of the wave-function is position-independent and then $\nabla\theta=0$,
one can obtain $Q/|W| =\frac{1}{2}$, which is the classical limit. 

However, when we consider the contribution from the 
$\beta$ terms, $Q/|W|$ will receive corrections as follows, 
\begin{eqnarray}
{\ddot{I}}_{\nabla\theta=0} 
& =& W   + 2Q + \int d\textbf{r} \frac{8\hbar^2\beta}{m} (\nabla^2\sqrt{\frac{\rho}{m}})^2  
\nonumber \\ && +\int d\textbf{r}2\hbar^2\beta m (\textbf{r}\cdot\nabla\Phi )\sqrt{\frac{\rho}{m}} \nabla^2\sqrt{\frac{\rho}{m}}\nonumber  \\
& & + \int d\textbf{r}\psi^* 4\hbar^2\beta m(\textbf{r}\cdot  \nabla)(\nabla\Phi\cdot\nabla\psi).
\end{eqnarray}
Since $(\nabla^2\sqrt{\frac{\rho}{m}})^2 \geq 0 $, it is also possible to obtain 
$Q/|W| \leq \frac{1}{2}$ for the position-independent wave-function. 
Thus, we have been able to study the effect of generalised uncertainty 
principle on quantum mechanical virial theorem. In this paper, we proposed that
GUP deformation can be important for gravitational fields, if the GUP parameter varies between 
different field theories. As the gravitational field has not been measured at
very short distances, such a deformation can be detected in near future experiments. 
We have analysed the effect of such a deformation on quantum mechanical 
virial theorem \cite{witten}, and demonstrated that this deformation
will have interesting consequences 
for such a quantum mechanical virial theorem.

It may be noted various other forms  of GUP have been proposed using a variety of physical motivations.  A different  deformation of the Heisenberg algebra 
has also been motivated from the  doubly special relativity 
 \cite{2}-\cite{3}. The doubly special relativity is a theory in which besides the speed of light, Planck energy is also a universal constant.
 Doubly special relativity has in turn been  motivated from  deformations of the 
  energy-momentum dispersion relation, and such  deformations occur in various approaches to quantum gravity, such as the 
  discrete space-time \cite{1q}, 
the spontaneous symmetry breaking of
Lorentz invariance in string field theory \cite{2q}, space-time foam models \cite{3q}, spin-network in loop quantum gravity \cite{4q}, 
non-commutative geometry \cite{5q}, and Horava-Lifshitz gravity \cite{6q}-\cite{6p}. It is possible to combine the deformation of the 
  Heisenberg algebra which is dependent on quadratic powers of momentum
  with the deformation 
  of the Heisenberg algebra produced by the  doubly special relativity  \cite{n4}-\cite{n5}. 
  The coordinate representation of the momentum operator for such a deformed Heisenberg algebra which is produced by the combination 
  of both these deformations contains linear powers of the momentum operator. This approach results in non-local fractional derivative contributions in any 
  dimension beyond the simple 
  one dimensional case. However, these fractional derivative contributions can be analysed  using the harmonic extension of function \cite{mir}-\cite{path}.  
 Another deformation of Heisenberg algebra containing inverse powers of momentum operator has also been proposed \cite{local}. 
 This deformation of the Heisenberg algebra is motivated by non-local quantum mechanics. It will be
  interesting to study the effects of these different forms of deformations of the Heisenberg algebra on small scale quantum systems with a gravitational potential.

It may be noted that the GUP deformation of quantum mechanics contains higher 
derivative terms, and such therms can produce   Ostrogradsky ghost \cite{aq}-\cite{bq}. 
These ghost states can in turn make the theory  non-unitary, and such theories can even contain negative norm states.  
However, it is possible to use different formalism to analyse such higher derivative states. It is possible to remove the negative norm states by  tracing  over a certain field configuration in the final state   \cite{hw}-\cite{hwhw}. Thus, it is possible to get rid of the negative norm states from such a higher derivative theory, but the theory would still  be 
  non-unitary. It is possible to use the Lee-Wick boundary conditions and obtain a  unitary theory \cite{w1}-\cite{lw}.  Thus, it is possible to analyse this deformed system using  an Lee-Wick field  corresponding  to a more massive mode.  Then  Lee-Wick boundary conditions can be imposed, and this could make the theory unitary. We would like to point out that such boundary conditions   violate causality  at microscopic scales, but  such a violation does not occur at  macroscopic  scale  \cite{w1}-\cite{lw}. The relation between Lee-Wick field theories,   nonanticommutative deformation of supersymmetric and GUP deformation has already been studied \cite{wl}- \cite{nona1}. Thus, it is interesting to analyse this system using Lee-Wick boundary conditions. 
It is interesting to note that   negative norm states  occur in usual gauge theories, but such negative states are removed using the Kugo-Ojima criterion which is  based on the   BRST symmetry  \cite{brst}- \cite{brst1}.  Such a  criterion  has also been used to remove the negative norm states in higher derivative theories \cite{r1}-\cite{j5}. In fact, various  superselection rules have been proposed to remove these ghost states from higher derivative theories 
\cite{b1}-\cite{b2}. 
Even though we do not address such questions in this paper, 
it is important to note that such higher derivative terms for 
GUP deformed systems can be consistently analysed. It would be 
interesting to analyse such terms using some of these approaches. 
It may be noted that usually such terms are analysed 
  in the framework of effective field theory, 
and such higher derivative 
terms occur \cite{effe}-\cite{effe1} in the derivative expansion of effective field theory. This is because these theories have a minimum 
measurable length scale associated with them, and so this is the scale that has to be integrated 
out in effective field theories \cite{ffn}. 
However, it would be also interesting to analyse these higher 
derivative terms using other approaches.

\section*{Acknowledgements}
 The work of Q.Z. is supported by NUS Tier 1 FRC Grant R-144-000-360-112.

\end{document}